\documentclass[12pt]{article}
\usepackage{epsfig}
\usepackage{cite}
\begin{document}

\hfill {\it Astropart. Phys.}, submitted\\

\begin{center}
{\large \bf Radiation fields of disk, BLR and torus in quasars and blazars:\\ implications for gamma-ray absorption}\\[1cm] Alina-C. Donea\footnote{Corresponding Author: adonea@physics.adelaide.edu.au, fax: +61 8 8303 4380} and R.J. Protheroe\\
Department of Physics and Mathematical Physics\\
The University of Adelaide, Adelaide, SA 5005, Australia
\end{center}

\begin{abstract}
The radiation fields external to the jets and originating from within a
few parsecs from the black hole, are discussed in this paper. They are
the direct radiation from an accretion disk in symbiosis with jets,
the radiation field from the broad line region (BLR) surrounding the
accretion disk, and the infrared radiation from a dusty torus.  The
jet/disk symbiosis modifies the energetics in the central parsec of
AGN such that for a given accretion rate, a powerful jet would occur
at the expense of the disk luminosity, and consequently the disk would
less efficiently ionize the BRL clouds or heat the dust in the torus,
thereby affecting potentially important target photon fields for
interactions of gamma-rays, accelerated electrons and protons along
the jet.

Motivated by unification schemes of active galactic nuclei, we briefly
review the evidence for the existence of broad line regions and
small-scale dust tori in BL~Lacs and Fanaroff-Riley Class I (FR-I)
radio galaxies.  We propose that an existing jet-accretion disk
symbiosis can be extrapolated to provide a large scale-symbiosis
between other important dusty constituents of the blazar/FR-I family.
In the present paper, we discuss in the context of this symbiosis
interactions of GeV and TeV gamma-rays produced in the jet with the
various radiation fields external to the jet in quasars and blazars,
taking account the anisotropy of the radiation.

\end{abstract}

\vspace{1em}


98.54.Cm  Active and peculiar galaxies

98.70.Rz  gamma-ray sources

\vspace{1em}


\newpage

\section{Introduction}

Following the introduction of unification schemes for active
galactic nuclei (AGN), it is now widely accepted that the central
nucleus of an active galaxy has a Kerr black hole, a relativistic
accretion disk, jets, a dusty torus and clouds emitting broad
emission lines.  Jets and accretion disks are strongly related
through a symbiosis, and we postulate that there may also be a
large scale symbiosis involving dust components.  Since
$\gamma$-rays are attenuated by photon-photon pair production
collisions with lower energy photons from the disk, BLR, or
torus, such a symbiosis could have an effect on the production
and absorption of $\gamma$-rays in AGN as the BLR and torus are
powered by the disk, and so may depend on the way in which the
jets are fed by the accretion disk.

A $\gamma$-ray flare could occur, for example, when a significant
amount of energy has been accumulated in the inner part of the
accretion disk and is subsequently expelled into the jets.  As a
consequence, relativistic shocks may form in the jets and could
account for the observed rapid and high amplitude $\gamma$-ray
variability.  The intense emission observed during a flare could
be also due to an increase in the accretion rate, in the bulk
Lorentz factor of the emitting material in the jet, or in the
efficiency of particle acceleration.  The production of
$\gamma$-rays from relativistic jets depends also on the density
of radiation fields since these photons can be scattered by
relativistic electrons to GeV and TeV energies, the external
radiation field giving additional target photons to the internal
synchrotron radiation assumed in the synchrotron self-Compton
(SSC) process \cite{Dermer93,Sikora94}. Mannheim
\cite{Mannheim95} and Protheroe \cite{Protheroe96} have shown
that the disk radiation can be relevant for $\gamma$-ray
production if the main constituents of the jet are relativistic
protons.

In this paper we shall concentrate on the role played by external
radiation fields in the absorption of $\gamma$-rays by
photon-photon pair production, and we shall calculate the pair
production opacity for GeV--TeV $\gamma$-rays emitted in the jets
of quasars and blazars.  We will only consider external radiation
fields existing in AGN, and will not discuss pair production by
$\gamma$-rays interacting with internal radiation fields in the
emission region as this has been discussed elsewhere
\cite{Ghisellini96,Bednarek99}.  We shall also not address the
problem of absorption by the infrared background radiation during
propagation to Earth
(e.g. refs.~\cite{Stecker98,ProtheroeMeyer00}).

We shall assume that the disk radiation, the BLR
radiation, and the torus radiation is anisotropic, but is
symmetric about the jet axis.  In such a radiation field the
reciprocal of the mean interaction length for photon-photon pair
production by $\gamma$-rays traveling along the jet axis is given
by
\begin{equation}
x^{-1}_{\gamma \gamma}(E,z)=\int_{\varepsilon _{\rm
min}}^{\infty} d \varepsilon \int_{\cos (\theta_{\rm max})}^{\cos
(\theta_{\rm min})} \frac{dn}{d \Omega}(\varepsilon, \theta ,z)(1
- \cos \theta) \sigma_{\gamma \gamma}(s) 2\pi \; d \cos \theta
\end{equation}
where $\theta$ is the angle between the direction of the soft
photons and the jet axis, $ \sigma_{\gamma \gamma}(s)$ is the
photon-photon cross section for pair production at center of
momentum frame energy squared $s=2E\epsilon(1-\cos\theta)$, $dn/d
\Omega$ is the differential photon number density, and $\cos
(\theta_{\rm max})$ and $\cos (\theta_{\rm min})$ are defined by
the disk, BLR, or torus geometry involved or, in the case of
$\cos (\theta_{\rm min})$, by the pair production threshold
condition. We calculate the optical depth for absorption of GeV
and TeV photons from the point of $\gamma$-ray emission to
infinity, in the relevant external radiation fields.

In quasars the torus and the BLR are active components, as can be
inferred from the large forest of emission lines observed, and
from the thermal infrared spectrum, allowing good estimation of
the diffuse radiation within the central parsec. In contrast,
blazars, which comprise BL Lacs and Optically Violently Variable
(OVV) quasars, are a special category of AGN having the jet
aligned closely to the line of sight.  The lack of emission lines
and UV bump in blazars is usually assumed to be due to a low
thermal activity at the centre, implying that the energy density
of the external radiation is small in blazars compared to that in
quasars.  We shall investigate whether or not this is the case,
and discuss how the $\gamma$-ray output could be different in
quasars and blazars depending on the dominant radiation field.

In Section 2 we shall discuss the optical/UV emission from an
accretion disk in symbiosis with the jet, and its effect on the
BLR emission clouds (ionized by disk photons) and the torus
(heating of dust).  It is difficult to estimate the disk
luminosity in blazars as has been done for some quasars with
visible UV bumps \cite{Donea96,Malkan82}.  Blazars do not
show thermal disk emission, and their spectra from radio to UV
frequencies, sometimes to X-ray frequencies, are universally
attributed to synchrotron emission from the relativistic jet. The
nonthermal boosted radiation from the jet dwarfs any thermal
emission, making the direct detection of BLR or torus activity,
and hence direct measurements of external photon fields present
within the parsec scale of blazars, impossible.

We assume that the accretion disk surrounding the black hole and
the relativistic jets are related by a small-scale symbiosis
\cite{Falcke95,Donea96}, and that the energy available for the
jet is the energy which would be dissipated between the last
marginally stable circular orbit (in Kerr geometry), $R_{\rm
ms}\approx 1.23 R_g$ where $R_g=GM/c^2$ is the gravitational
radius, and the radius of the base of the jet $R_{\rm jet}>R_{\rm
ms}$. The total power of the jet depends on $R_{\rm jet}$.  A
thick jet base means that more energy is expelled into the jet,
and therefore, less energy is dissipated in the disk.  A direct
consequence of this assumption is that an accretion disk in
symbiosis with jets (ADJ) produces a disk luminosity smaller than
that of a ``standard'' \cite{Novikov73} relativistic accretion
disk.  We suggest that the ADJ model could describe
low-luminosity accretion disks in some blazars, and may provide a
bridge between the standard accretion disk assumed to exist in
quasars and Fanaroff-Riley Class II (FR-II) sources
\cite{Reynolds96} and the low-power advection dominated accretion
disk which may be found in some FR-I sources.

In Section 3, assuming reasonable models for the size and
distribution of BLR clouds, we calculate the energy density of
the radiation field of the BLR, and the optical depth
for absorption of GeV and TeV photons. Broad Line Regions
surround accretion disks and may also be important sources of
soft photons which could be inverse-Compton scattered by
relativistic electrons in the jet to produce $\gamma$-rays in
``external Compton models'' \cite{Sikora94}.

In section 4, we discuss the infrared radiation from a dusty
torus.  Based on the assertion that FR-I and BL Lacs are related
\cite{UrryPadovani95}, and that there is strong evidence that
some FR-I have dusty tori \cite{Antonucci01}, we assume that
blazars must also have dusty tori \cite{DoneaProtheroe02a} and
discuss what evidence there is to support this.  Since the
infrared emission produced by tori could inhibit the escape of
TeV $\gamma$-rays due to photon-photon pair-production
\cite{ProtheroeBiermann97} we study how changes in disk activity
can modify the geometry and the heating mechanism of the torus,
and how this affects the optical depth for $\gamma$-$\gamma$
absorption in the torus infrared radiation.


\section{Radiation  from an accretion disk with jets}

Relativistic outflows are very nicely described by a model of the
small-scale symbiosis between the accretion disk and jets
\cite{Falcke95}.  Donea and Biermann \cite{Donea96} have used
such a model for an accretion disk with jets starting at the
inner region of a disk, and were able to reproduce the UV bump in
quasars. By fitting observed spectra they derived upper limits
for the radius of the base of the jet, $R_{\rm jet}$, and they
found that $R_{\rm jet}$ cannot be too far from $R_{\rm ms}$.
The jet base, or ``footring'', is actually the thin layer between
$R_{\rm ms}$ and radius $R_{\rm jet}$, the jet being approximated
by a hollow cylinder of inner radius $R_{\rm ms}$ and outer
radius $R_{\rm jet}$, such that $R_{\rm ms} \le R_{\rm jet} \ll
R_{\rm out}$, where $R_{\rm out}$ is the outer radius of the
disk.

The ADJ model assumes that the gravitational potential energy
available between $R_{\rm ms}$ and $R_{\rm jet}$ is the energy
reservoir of the jets.  The total power of the jets is strongly
dependent on the accretion mass rate in the disk and on the size
of its footring (we not include here the interaction between the
black hole and jets which could also result in it putting a
non-negligible fraction of its energy into the jets
\cite{Blandford77}).  The total power of the jets is
\begin{equation}
Q_{\rm jet} = L_{\rm disk} - L_{\rm disk}^{\rm jet}
\end{equation}
where $L_{\rm disk}^{\rm jet} $ is the luminosity of the disk
with jets and $L_{\rm disk} $ would be the luminosity of the disk
if the conditions necessary to drive outflows were not met, i.e.,
the case of a standard relativistic disk.

A coupled jet-disk system must obey the laws of conservation of
mass and angular momentum.  We assume that the jet is fed with
mass by the accretion disk, and that the flow of mass into the
jet per unit of time $\dot{M}_{\rm jet}=q_{ m}\dot{M}$ is a
fraction $q_{ m} \le 1$ of the accretion mass rate into the disk
$\dot{M}=-2 \pi R \Sigma u^R$, where $\Sigma$ is the surface
density of mass in the disk and $u^R$ is the radial velocity of
gas at given radius $R$.  The equation for conservation of mass
requires $\dot {M}=\dot {M}_{\rm disk}+\dot {M}_{\rm jet}$.
Derivation of the relevant equations can be found in Donea and
Biermann \cite{Donea96}, who followed the standard method of
calculating the emission spectrum from an accretion disk
\cite{Novikov73}.  If there is no angular momentum and mass loss
into the jet the equations used for mass and angular momentum
transport in the disk become the standard equations
\cite{Novikov73}, with $R_{\rm jet} \to R_{\rm ms}$, $ Q_{\rm
jet} \to 0$ and $L_{\rm disk}^{\rm jet} \to L_{\rm disk}$.  In
this paper, we shall adopt $q_m=0.1$ as used by Falcke et
al.~\cite{FalckeMalkanBiermann95} in interpreting the radio-UV
correlation in AGN.

The local physics at the inner radius of the disk (radius $R_{\rm
jet}$) is directly related to the extraction of angular momentum
from the in-falling gas, and so modifies the structure of the
relativistic disk \cite{DoneaBiermann01}.  From this one
calculates the dissipation energy at radius $R$, $D^*(R)$, which
must be done numerically in the case of a Kerr black hole, and a
detailed discussion of this would serve no useful purpose here.
Instead, for the purpose of illustration, and for the sake of
simplicity we give here only the relation for the simpler case of
a Schwarzschild black hole:
\begin{equation}
D^*(R) = \frac{3 G M \dot M}{8 \pi R^3} \bigg[ (1-q_m) - (1-q_m)
{\bigg( \frac {R_{\rm jet}}{R} \bigg) }^{1/2} \bigg]
\end{equation}
where $q_m = \dot{M}_{\rm jet}/\dot{M}$ and $M$ is the black hole
mass; nevertheless, in the present paper we use $D^*(R)$ for the
disk of a Kerr black hole.  A reasonable outer radius of the
footring of the jet would be $R_{\rm jet} \le 10 R_{\rm
g}\approx 10^{-4} M_8$~pc where $M_8=M/10^8 M_{\rm \odot}$.

\begin{figure}
\begin{center}
\epsfig{file=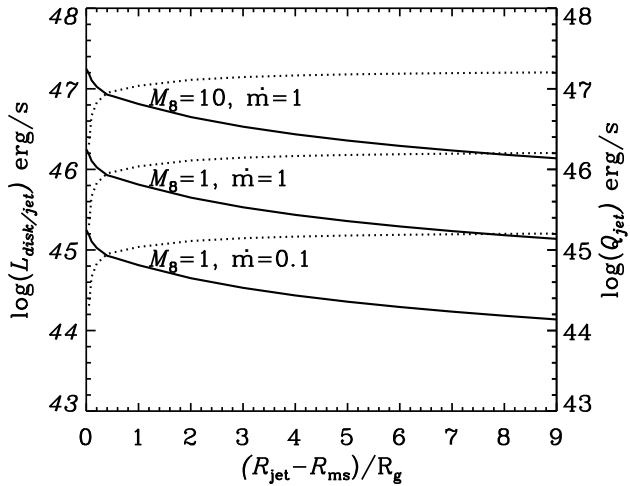,width=10.0cm}
\caption{Luminosities of disks with jets vs.\ the thickness of
the jet's footring $(R_{\rm jet} - R_{\rm ms})$ (solid curves).
The dotted lines show $Q_{\rm jet}$ vs.\  the thickness of the
footring. ($M_8\equiv M/10^8 M_{\odot}$, $q_{m}=0.1$). }
\label{fig1}
\end{center}
\end{figure}

The disk/jet symbiosis is reflected mainly in a modified photon
spectrum from the inner region of the accretion disk where the jet is
anchored. The important result is that the spectrum from a Kerr
accretion disk is cut off at high frequencies, from extreme UV to soft
X-rays.  In Fig.~\ref{fig1} we plot disk luminosities for the ADJ
model, versus the thickness of the footring of the jet for Kerr black
holes with masses $M=10^8 M_{\odot}$ and $M=10^9 M_{\odot}$, and for
different mass accretion rates given by $\dot m=\dot M/\dot M_{\rm
edd}$, where $M_{\rm edd}$ is the Eddington accretion rate.  The case
$R_{\rm jet}=R_{\rm ms}$ corresponds to an accretion disk without jets
(``standard'' accretion disk) and $Q_{\rm jet} =0$. As can be seen,
for thicker jet bases more energy is available to power jets (dotted
curves show $Q_{\rm jet}$ increasing with $R_{\rm jet}-R_{\rm ms}$)
and less energy is radiated by the disk (solid curves show $L_{\rm
disk}$ versus thickness).  We note that a small variation of the
geometry at the coupling between jet and disk at radii $R_{\rm jet}$
less than $\sim 3 R_g$ would induce large variations in the power of
the jet, possibly causing flare activity in blazars. This is because,
in a Kerr metric, at small radii close to the black hole there is a
large amount of gravitational potential energy available for
dissipation into the jet.

Because of the jet/disk coupling, the ADJ model predicts a lower
disk luminosity and a softer photon spectrum, and some of the
implications of this for interactions of accelerated electrons or
protons are discussed in a separate paper
\cite{DoneaProtheroe02b}.  As mentioned earlier, a flaring state
could arise from enhanced AGN central activity whereby the jets
get more energy from the disk.  In this case, the disk changes
from an ADJ with weak jets into an ADJ with a spectrum cut off at
high frequencies but with correspondingly stronger jets.  Since,
in this case, the average energy of photons from an ADJ with
strong jets (flaring) becomes lower than the average energy of
photons from an ADJ with weak jets this will change the shape of
the resulting $\gamma$-ray spectrum in
models~\cite{Sikora94,DoneaMasnou01} where disk photons are
inverse-Compton scattered by relativistic electrons in the jet.
 
Luminosities of disks in blazars are probably  $L_{\rm disk}
< 10^{46}$~erg/s, while quasars typically have $L_{\rm
disk}=10^{46}$--$10^{48}$~erg/s.  The ADJ model gives a
simplified approach to the symbiosis between the disk and jets,
and could describe low-luminosity accretion disks for quasars and
some blazars.  However, for some blazars the ADJ model could
still give UV fluxes much higher than those observed, and in this
case an ejection dominated accretion flow (EDAF) model, which is
an ADJ model with a wind \cite{DoneaFalckeBiermann99}, may be
better. An EDAF model has been applied to the central activity of
Sgr A* where it was shown that a wind plus jets extracts energy
from the disk more efficiency. Since the EDAF model does not
leave much energy to be dissipated in the accretion disk, the
radiation field from the disk would become even less important
for GeV and TeV $\gamma$-ray absorption.  Alternatively, the jets
could extract more energy from the accreting gas if the disk with
a jets turns into an advection dominated accretion flow 
(ADAF) \cite{Narayan94}.

\begin{figure}
\begin{center}
\centerline{\epsfig{file=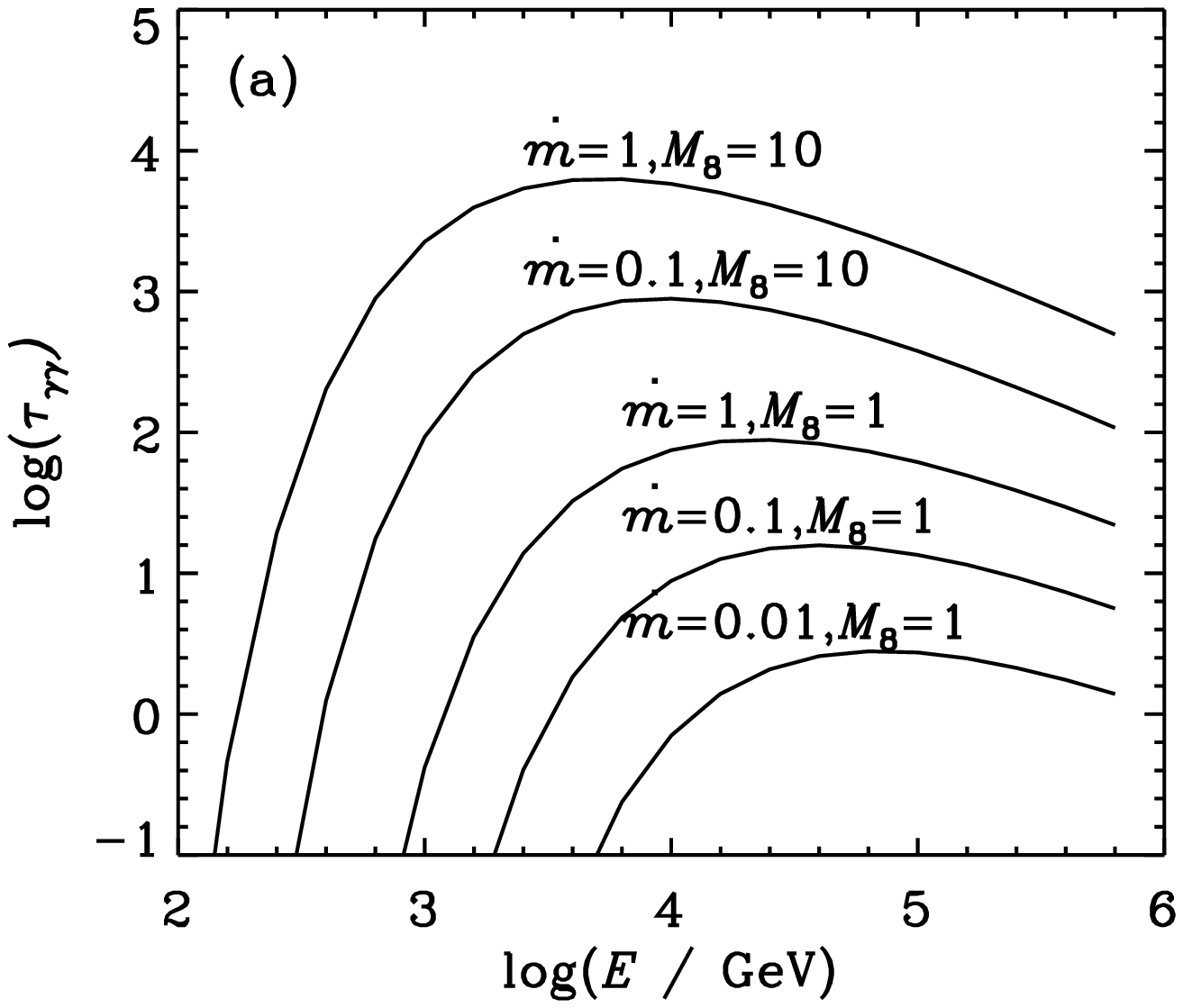,width=10.0cm}\hspace*{-5em}\epsfig{file=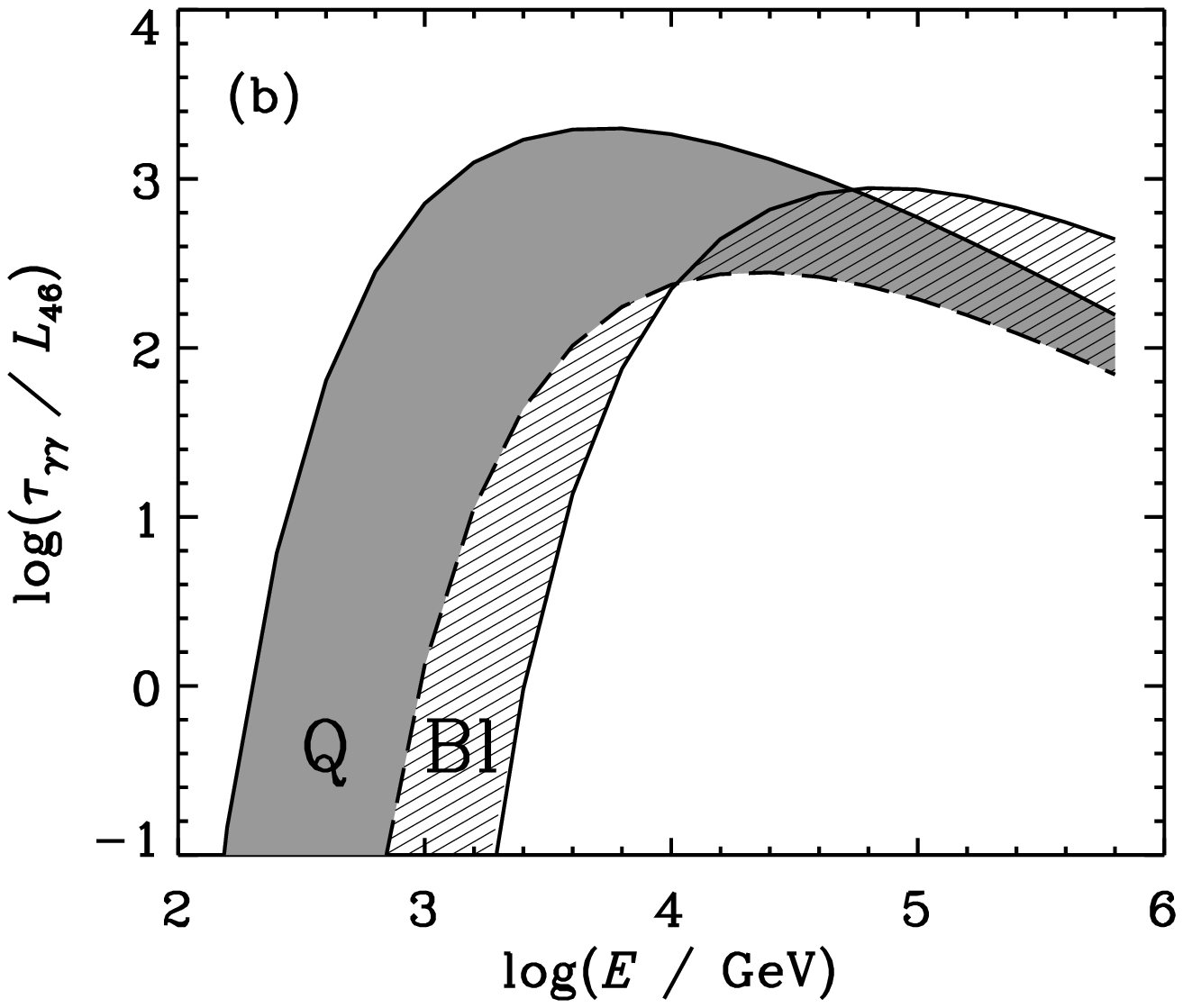,width=10.0cm}}
\caption{Optical depths from $z=0.01$~pc to infinity for
$\gamma$-rays produced in the jet interacting with UV disk
photons for the case of an ADJ model with $q_m=0.1$ and $R_{\rm
jet}=10R_g$: (a) for various black hole masses and accretion
rates indicated by the labels attached to the curves; (b) optical
depth normalized to disk luminosity in units of $10^{46}$ erg
s$^{-1}$ showing the probable range of $\tau_{\gamma\gamma}$ for
blazars ``Bl'' and quasars ``Q''.}
\label{fig10}
\end{center}
\end{figure}

Fig.~\ref{fig10}(a) shows the optical depths from $z=0.01$~pc to
infinity for absorption of GeV-TeV photons produced in the jet
and interacting with photons from an ADJ with $q_m=0.1$.  We see
that at 1~TeV photons are only absorbed for high black hole
masses and mass accretion rates close to the Eddington limit.
This could apply to quasars where the disk radiation field has a
high density around the base of the jet.  For low accretion mass
rates with $\dot M \le 0.1 M_{\rm edd}$, and lower black hole
masses, $M \le 10^8M_\odot$, the photon-photon optical depth is
$\tau_{\gamma \gamma} <1$ at 1~TeV.  We note that the disk
luminosity corresponding to the case $\dot{m}=1$ for $M_8=1$ lies
roughly at the boundary between blazars and quasars.  In
Fig.~\ref{fig10}(b) we plot the optical depth divided by
$L_{46} \equiv L_{\rm UV}/10^{46}$~erg s$^{-1}$, and give the likely
range for blazars and quasars using $\dot{m}=1$ for $M_8=1$ as
the boundary.  In blazars we may have ADJ models with accretion
rates as low as $\dot{m}=0.01$--0.001 -- at lower accretion rates
an ADAF or EDAF model including jets may be more realistic.  We
use $M_8=1$ and $\dot{m}=0.01$ as the low luminosity boundary of
blazars with an ADJ in Fig.~\ref{fig10}(b).  Similarly, we use a
supermassive black hole ($M_8=10$) and accretion close to the
Eddington limit ($\dot{m}=1$) for the upper luminosity boundary
of quasars in Fig.~\ref{fig10}(b).  Of course, these boundaries
are by no means rigid and there will be some overlap between the
two populations, particularly when we plot
$\tau_{\gamma\gamma}/L_{46}$.

The ADAF model seems to explain
better very low luminosity blazars, but it cannot energize the
BLR or heat efficiently dust in the torus; the lack of observed
BLR and torus activity for some blazars has been sometimes
interpreted as suggesting that blazars do not have tori or
BLR. We consider that if blazars are to be included in the
unification schemes they should also have BLR and dusty tori, but
possibly with lower emission than in quasars, and we shall
discuss this possibility later.

\section{Photons from broad line region clouds}

We are interested in obtaining the optical depth for absorption
of GeV and TeV photons in the anisotropic radiation field of the
BLR. For this we need to know the distribution of the BLR clouds
around the jet axis. Broad line regions in AGN are not spatially
resolved, but information about their size and the gas
distribution can be obtained by reverberation mapping, from which
it appears that BLR are stratified, and have typical sizes of the
order of light-weeks in Seyfert galaxies and light-months in
quasars \cite{Peterson94}.

Despite the fact that the majority of BL Lacs do not show any
emission lines, there are observations suggesting that there is a
weak broad $H_{\alpha}$ emission line in BL Lacertae itself
\cite{Corbett00}, and even Mrk 421 seems to have a very low BLR
luminosity, $L_{\rm BLR} = 1.5 \times 10^{40}$~erg/s
\cite{Morganti92}.  Interestingly, Celotti et
al.~\cite{Celotti98} have suggested, based on the rapid TeV
variability of Mrk 421, that its accretion rate is lower than
$10^{-2}-10^{-3}$ of the Eddington rate. Hence, a less efficient
ADJ or an ADAF may then power the BLR for this source. For Mrk
501, the BLR luminosity has been found to be $L_{\rm BLR} = 4
\times 10^{41}$~erg/s \cite{Stitckel93}. In addition, Chiaberge et
al. \cite{Chiaberge00} have concluded that BLR and obscuring tori
are closely linked in quasars, and both are present only in
association with accretion.

In order to model the BLR, and hence its radiation field, we need to
know its size and the emission line emissivity as a function of
radius.  The two-phase (hot/cold) model of the BLR predicts a
relatively thin shell-like BLR \cite{Krolik81}.  Higher
ionization lines are found at smaller radii (e.g. for NGC 5548
\cite{Baldwin95}), and this may favour a thin shell model for the
BLR.  This provides a very simple model of the BLR in which the total
BLR region luminosity $L_{\rm BLR}$ is emitted from the surface of a
sphere of radius $r_{\rm BLR}$, i.e. $L_{\rm BLR}/4\pi r_{\rm BLR}^2$
per unit area.

Although it is not known if strong winds can exist in blazars, there
are models which explain the formation of clouds from bloated stars
exposed to winds from accretion disks \cite{Alexander99,Murray95}, and shocks from jets could also contribute to
cloud formation \cite{Ghisellini96}. On the other hand, Frometh and Melia
\cite{Fromerth01} suggest that the unsteady, turbulent
accretion flow onto the black hole is subjected to several
disturbances capable of producing shocks, which would heat the gas and
allow for formation of BLR clouds. They also suggest that circular
clouds originate at smaller radii.  Results from reverberation mapping
\cite{Peterson94} suggest that the BLR in Seyferts is extended
with an outer radius at least 10 times larger than the inner radius,
and Kaspi and Netzer \cite{Kaspi99} use models with a radial dependence of BLR
cloud number density and radius to fit their data.

Clearly, there are several possible mechanisms for formation of BLR
clouds and it is not known which one operates in blazars.  It could be
that all these mechanisms play a role in BLR cloud formation.  We
shall adopt two extreme cases for the distribution of BLR cloud
emissivity: (i) a geometrically thin shell at radius $r_{\rm BLR}$,
(ii) a geometrically thick shell extending from radius $r_{\rm
in,BLR}$ to $r_{\rm out, BLR}$ with the radial dependence of line
emissivity based on the best fit parameters of Kaspi and Netzer \cite{Kaspi99}
used in their modelling of the BLR in NGC 5548 (number density of
clouds $n_{\rm cl} \propto r^{-1.5}$, cloud radius $R_{\rm cl} \propto
r^{0.6}$).  In both cases we shall assume that the line emission is
isotropic, and that the ionization source (the accretion disk) is
located at $r=0$ and has luminosity $L_{\rm UV}$.  Accretion disk
photons excite the BLR material, and emission lines such as
$H_{\alpha}$, $H_{\beta}$, $OIII$, $NII$, etc., are produced. We
simplify our problem by making the approximation that the entire BLR
luminosity is emitted in the $H_{\alpha}$ (6563 \AA) line.

\begin{figure}
\begin{center}
\epsfig{file=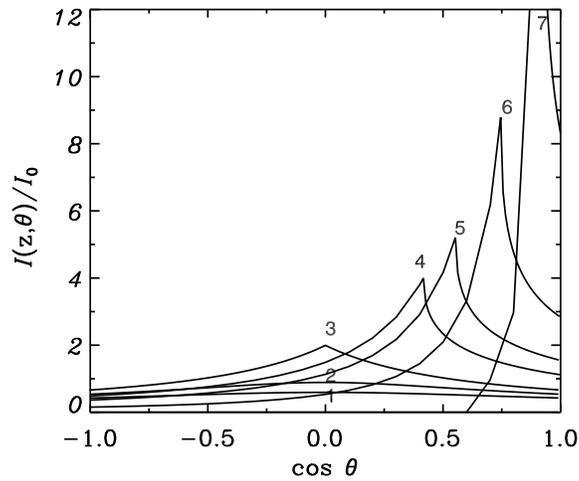,width=10.0cm} 
\caption{The angular distribution of the BLR $H_{\alpha}$
emission line intensity $I(z,\theta)$ for different distances
from to the centre along the jet. The BLR is modeled as a
geometrically thin shell at radius $0.01$~pc. Labels 1--7
correspond to $z=0, 0.008, 0.009, 0.01, 0.011, 0.012, 0.015$~pc,
respectively. The anisotropy increases when the blob moves away
from the BLR. The normalization factor $I_{0}$ is given by
$I_0=cu_0/4 \pi $, with $u_0=L_{\rm BLR}/(4 \pi z^2 c$).}
\label{fig5}
\end{center}
\end{figure}

The BLR is assumed to be optically thin with optical depth
$\tau_{\rm BLR}$ such that the total BLR luminosity is $L_{\rm
BLR} \approx \tau_{\rm BLR}L_{\rm UV}$.  Celotti et al.\
\cite{Celotti97} analyzed a sample of quasars and found an
average value of $L_{\rm BLR} \approx 4 \times 10^{45}$~erg/s,
and for a sample of 12 blazars (including Mrk 421 and Mkr 501)
found an average value of $L_{\rm BLR} \approx 2 \times
10^{43}$~erg/s. However, flat spectrum radio quasars and OVV
quasars could have higher $L_{\rm BLR}$ as required by
Blazejowski et al.~\cite{Blazejowski00} in order to explain the
$\gamma$-ray spectra for these objects.  In our calculations we
use $L_{\rm UV}=10^{46}$ erg/s and $\tau_{\rm BLR}=0.01$, which
are typical values used for fitting the broad emission lines with
photoionization models (for $\tau=0.01$, $L_{\rm BLR}=
10^{44}$~erg/s).  For $\gamma$-ray attenuation by photon-photon
pair production we need the photon number density per solid angle
of BLR photons along the jet axis, $dn /
d\Omega(\theta,z)=I(\theta,z)/c$ where $I(\theta,z)$ is the
intensity.  Calculation of the intensity is straightforward for
distribution (i) where the BLR is a geometrically thin shell.  In
Fig.~\ref{fig5} we show the full angular dependence of the BLR
$H_\alpha$ intensity at different distances $z$ along the jet for
the case where the BLR is a thin shell at $r_{\rm BLR}=0.01$~pc.

\begin{figure}
\begin{center}
\epsfig{file=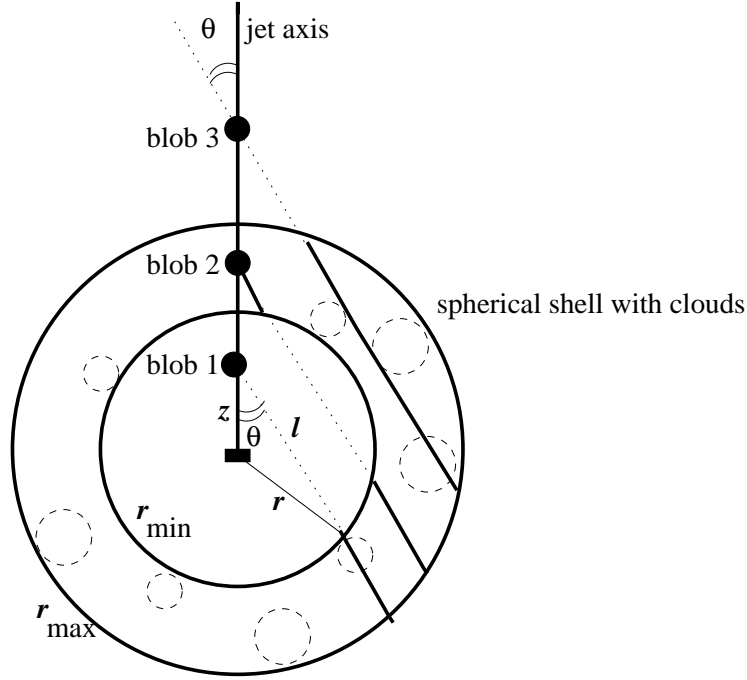,width=10.0cm} 
\caption{Sketch (not to scale) illustrating the geometry used for
computation of the anisotropic intensity of the BLR radiation for
distribution (ii) where the BLR is a geometrically thick
spherical shell. The $\gamma$-ray emitting regions travel
outwards along the jet axis and may be inside the cavity of the BLR
(blob 1), within the BLR (blob2) or outside the BLR (blob 3),
depending on their distance along the jet $z$.}
\label{fig4}
\end{center}
\end{figure}

For distribution (ii) we shall calculate the intensity by integrating
the emissivity along lines of sight through the BLR sketched in
Fig.~\ref{fig4}.  A fraction $d\tau_{\rm BLR} =n_{\rm cl} \sigma_{\rm
cl} dr$ of the central luminosity would then be re-processed in $r \to
(r+dr)$.  Since this spherical shell has volume $4 \pi r^2 dr$ the
emissivity (in erg s$^{-1}$ sr$^{-1}$ cm$^{-3}$) of re-processed
radiation inside the BLR at radius $r$ is
\begin{equation}
 j(r)=\frac{L d \tau_{\rm BLR}}{16 \pi^2 r^2 dr}=\frac{L n_{\rm
 cl}\sigma_{\rm cl}}{16 \pi^2 r^2 }.
\end{equation}
For $n_{\rm cl} = n_0 (r/r_{in, \rm BLR})^{\alpha}$ and $R_{\rm cl} =
R_0 (r/r_{in,\rm BLR})^{\beta}$, where $n_0$ and $R_0$ is the number
density and the radius of clouds at $r_{\rm in, BLR}$, respectively,
the optical depth of the entire BLR to central UV radiation is
\begin{equation}
\tau_{\rm BLR} = \int_{r_{\rm in, BLR}}^{r_{\rm out, BLR}} n_0
\pi R_0^2 \left( \frac{r}{r_{in,\rm BLR}} \right)^{\alpha+2
\beta} dr 
\end{equation}
from which we obtain the constant $n_0 R_0^2$ for any assumed
value of $\tau_{\rm BLR}$.  Note that we shall use the power-law
exponents preferred by Kaspi and Netzer \cite{Kaspi99},
$\alpha=-1$ and $\beta=-1.5$.  We treat separately the three
possible emission region locations shown in Fig.~\ref{fig4}: (a)
blob~1 inside the cavity inside the BLR, (b) blob~2 located
within the BLR region, and (c) blob~3 is outside the BLR.  The
intensity of radiation at angle $\theta$ to the jet axis at a
position $z$ {\em within} the BLR region (e.g.\ blob~2 in
Fig.~\ref{fig4}) is given by
\begin{equation} 
I(z,\theta)=\int_{0}^{l_{min1}} j(r) dl+\int_{l_{min2}}^{l_{\rm max}} j(r) dl,
\end{equation} 
where
\begin{equation}
r^2=z^2-2 z l \cos \theta + l^2, 
\end{equation}
\begin{equation}
l_{\rm max}=z \cos \theta + z \left( \cos ^2 \theta +
\left( \frac{r_{out, BLR}}{z} \right)^2 -1 \right)^{1/2}, 
\end{equation}
and
\begin{equation} 
 l_{\rm min 1,2}=z \cos \theta \pm z \left( \cos ^2 \theta + \left( \frac{r_{in, BLR}}{z} \right)^2 -1 \right)^{ 1/2}. 
\end{equation}

\begin{figure}
\begin{center}
\epsfig{file=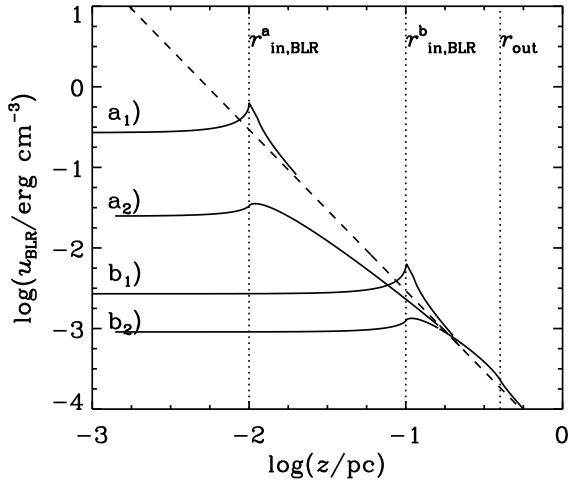,width=10.0cm} 
\caption{Energy density of radiation from BLR clouds at different
distances $z$ along the jet axis.  Cases labelled a$_1$ and a$_2$ are
calculated for a BLR with $r_{\rm BLR}=0.01$~pc or $r_{\rm in,
BLR}=0.01$~pc. Cases labelled b$_1$ and b$_2$ correspond to a BLR with
$r_{\rm BLR}=0.1$~pc or $r^b_{\rm in, BLR}=0.1$~pc. The label
subscripts indicate the BLR cloud distribution used: 1 --
geometrically thin spherical shell (distribution i) ; 2 --
geometrically thick spherical shell (distribution ii) with $r_{\rm
out, BLR}=0.4$~pc, $\alpha=-1$, and $\beta=-1.5$.  The dashed line
shows the crude approximation $u_{\rm BLR} \approx L_{\rm BLR}/4 \pi c
z^2$, with $L_{\rm BLR}=10^{44}$~erg/s.}
\label{fig6}
\end{center}
\end{figure}

The total (angle-integrated) energy density of BLR is:
\begin{equation} 
u(z)=\int \frac{I(z,\theta)}{c} d \Omega,
\end{equation}
and is plotted in Fig.~\ref{fig6} as a function of distance along the
jet.  The energy density of BLR photons clearly depends also on the
distribution of BLR clouds.  At $z \gg r_{ \rm BLR}$ (distribution i)
and $z \gg r_{ \rm out, BLR}$ (distribution ii) the BLR photon energy
density approaches the limit $u_{\rm BLR} \approx L_{\rm BLR}/4 \pi c
z^2$ (dashed line in Fig.~\ref{fig6}).  For distribution (i), where we
have a geometrically thin spherical shell for the BLR, the energy
density peaks sharply at $r_{\rm BLR}$ as found also by Ghisellini and
Madau \cite{Ghisellini96}.  In the case of distribution (ii) the energy density
peaks near $r_{\rm in, BLR}$, but the curves are smoother as a result
of the finite thickness of the BLR.  Both cases are idealizations and
there is probably a spread in cloud size at any given radius. From
Fig.~\ref{fig6} we see that it is very important to know the right
distribution of BLR clouds since there can be a difference of more
than two orders of magnitude between the energy density of photons
calculated for the same BLR luminosity but different BLR cloud
distributions.

\begin{figure}
\begin{center}
\centerline{\epsfig{file=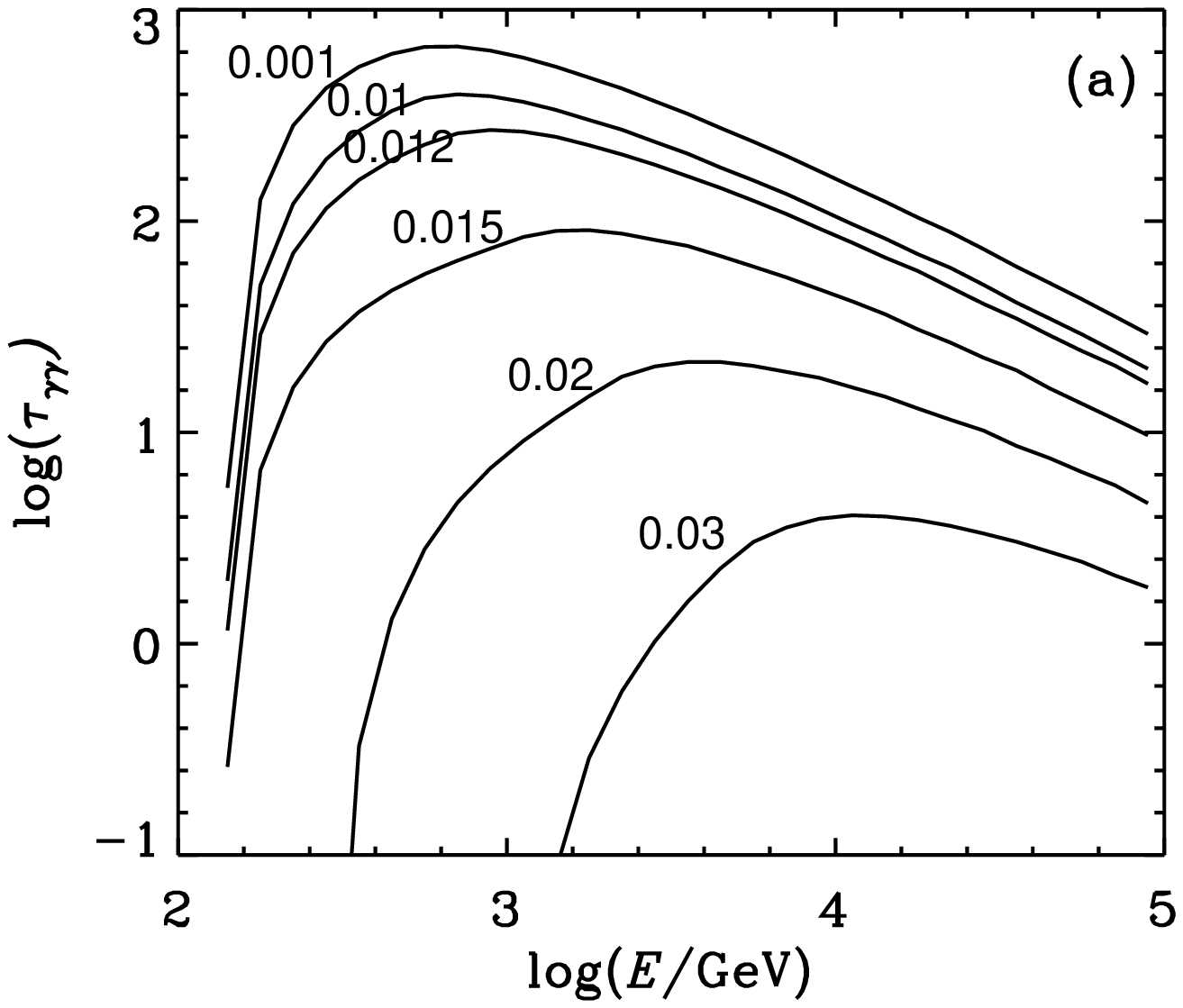,width=10.0cm}\hspace*{-5em}\epsfig{file=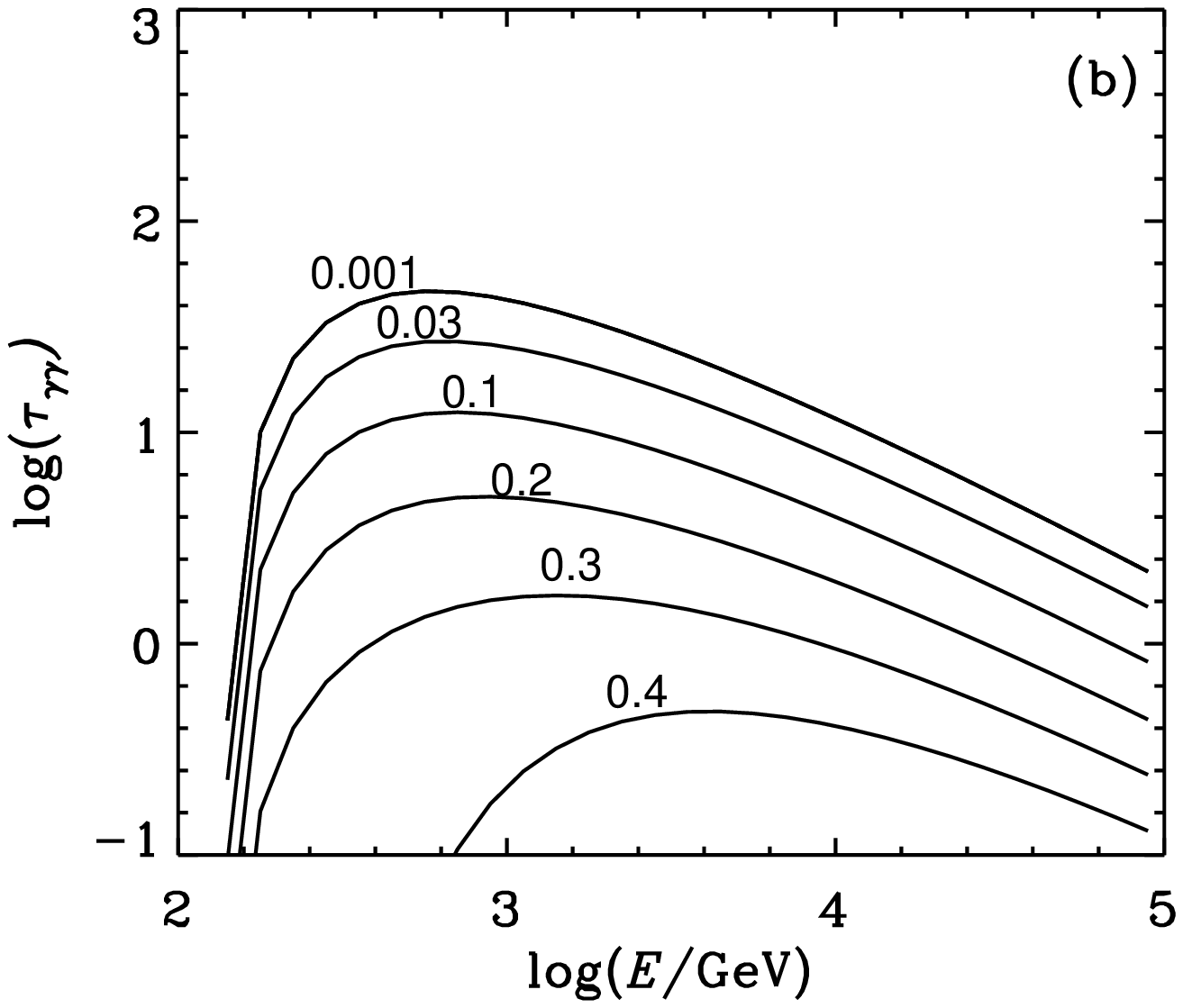,width=10.0cm} }
\caption{Photon-photon pair production optical depths for
$\gamma$-rays traveling along the jet axis interacting with
photons from a BLR having luminosity $L_{\rm BLR}=10^{44}$~erg/s
and comprising: (a) a geometrically thin spherical shell of
radius $r_{\rm BLR}=0.01$~pc; (b) a geometrically thick spherical
shell (distribution ii) with $r_{\rm in,BLR}=0.01$~pc, $r_{\rm
out, BLR}=0.4$~pc, $\alpha=-1$, and $\beta=-1.5$. Numbers
attached to the curves give $z_0$ in pc.}
\label{fig8}
\end{center}
\end{figure}

In Fig.~\ref{fig8}(a) we show the optical depth from various
emission region heights, $z_0$, to infinity for absorption of
100~GeV--100~TeV $\gamma$-rays in the radiation field of the BLR
for distribution (i) where the BLR clouds are located in a
geometrically thin shell at radius 0.01~pc. The
optical depth is plotted for a BLR luminosity of $10^{44}$~erg/s,
but can be linearly scaled for other BLR luminosities.  The
anisotropy of the BLR radiation field is reflected in the shapes
of the curves as one goes to higher $\gamma$-ray energies. 
Fig.~\ref{fig8}(b) gives the optical depth for cloud distribution
(ii), with the inner radius at 0.01~pc (the same as $r_{\rm BLR}$
in part a) and outer radius 0.4~pc.  In this case, while at
$z=0.01$~pc the optical depth is less than for distribution (i),
$\tau_{\gamma\gamma}$ drops off more slowly with $z_0$, such that at
only slightly larger distances the optical depth becomes orders
of magnitude higher than for distribution (i) until $r^b_{\rm
out, BLR}$ is reached.  This is in part because the possibility
of near head-on collisions of $\gamma$-rays with BLR photons for $z
<r^b_{\rm out, BLR}$ for distribution (ii) -- large angles
between directions of interacting photons means more collisions
will be above the pair production threshold.

\begin{figure}
\begin{center}
\epsfig{file=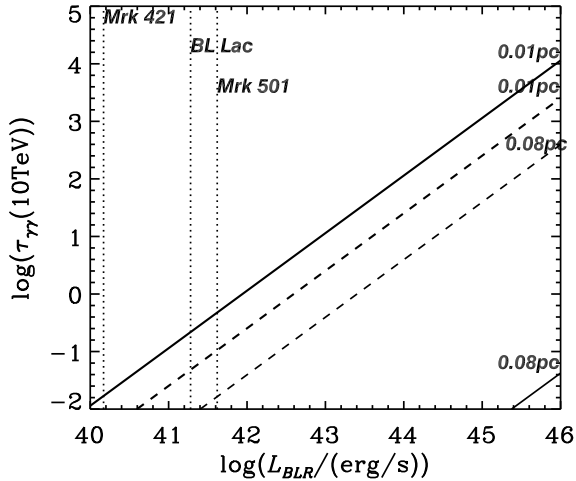,width=10.0cm} 
\caption{Photon-photon pair production optical depth for 10 TeV
photons from $z_0$ (numbers attached to the curves) to infinity for
$L_{\rm BLR} = 10^{44}$ erg/s.  Solid lines correspond to distribution
(i) with $r_{\rm BLR}=0.01$~pc, dashed to distribution (ii) with
$r_{\rm in,BLR}=0.01$~pc, $r_{\rm out, BLR}=0.4$~pc, $\alpha=-1$, and
$\beta=-1.5$. }
\label{fig18}
\end{center}
\end{figure}

Fig.~\ref{fig18} shows the $\gamma$-$\gamma$ opacity for 10 TeV
photons from $z=z_0$ to infinity vs.\ $L_{\rm BLR}$ for the two cases
shown in Fig.~\ref{fig8}.  For blazars with a very low BLR luminosity,
10 TeV photons can probably escape through the BLR. However, in
quasars BLR typically have $L_{\rm BLR} \ge 10^{44}$ erg/s, and so
absorption of 10 TeV $\gamma$-rays can be quite important if the emission
region is inside the BLR (distribution i) or is close to $r_{\rm
in,BLR}$ (distribution ii).

We comment here on a very intriguing problem, namely the absence
of a Ly absorption edge in quasars \cite{Maiolino01}. It was
expected that BLR clouds situated along the line of sight should
produce a sharp Ly absorption edge in more than 30\% of quasars
\cite{Antonucci89}. However, Maiolino et al.~\cite{Maiolino01}
have noted that none of the 101 quasars in the sample analyzed by
Zheng et al.~\cite{Zheng97} show this feature. There are two
possible explanations for the missing edge: (a) BLR clouds have a
sufficiently large component of their random velocity along our
line of sight to result in a broadening of the Ly edge and
thereby making it difficult to observe, or (b) the BLR does not
have a spherically symmetric distribution around the ionizing
accretion disk, such that there are no BLR clouds intersecting
our line of sight to the disk, and Maiolino et
al.~\cite{Maiolino01} suggest a flattened geometry could do this.
Such a flattening in the cloud distribution could produce weaker
densities of BLR photons and so would impede less the escape of
TeV photons from the central region.  However, we adopt a
spherical distribution of clouds, and instead interpret the
absence of an observed Ly absorption edge in 101 quasars to
suggest that the BLR clouds must have a covering factor $f_{\rm
cov}\sim 0.03$ or less (95\% upper limit; for a covering factor
0.01 the probability of a BLR cloud being in the line of sight to
the disk in none of 101 observations is 36\%, quite consistent
with the observations). For a given covering factor, $f_{\rm
cov}$, the fraction of the disk luminosity intercepted by the BLR
must then be less than $f_{\rm cov}$, and justifies our
use of $\tau_{\rm BLR}=0.01$.  Nevertheless, $\tau_{\rm BLR}
\sim 0.1$ have also been widely used, and would lead to
photon-photon pair production opacities a factor of 10 higher than
those presented here.

\section{Radiation from a dusty torus }

Chiaberge et al.~\cite{Chiaberge99} have suggested that tori do
not exist in FR-I sources, or they must have a very thin geometry.
This conclusion results from the correlation found between
optical and radio cores of some FR-I sources.  However, the
sample used in this analysis represents only just under half of
the FR-I galaxies tabled by Zirbel and Baum \cite{Zirbel95}, and
one needs additional observations of the polarized optical
spectrum to reach a firm conclusion about this
\cite{Whysong01}. Nevertheless, these results could be
interpreted as suggesting that perhaps there is a correlation
between a flat infrared torus and a flat distribution of BLR
clouds as suggested by Maiolino \cite{Maiolino01}.  We do not know
what mechanism, other than accretion, could cause such a
flattening of the BLR and the torus. However, such a flattening
might be understood in terms of a non-homogeneous spherical shell
of BLR or dust in which the polar caps above and below the disk
plane are much more diluted than the equatorial belt. Zier and
Biermann \cite{ZierBiermann01} have shown that the caps of a
spherical layer of dust could be diluted so much, that even a
dense torus would have a doughnut shape.  



We have already discussed some aspects related to the infrared
radiation from ubiquitous dusty tori in quasars.  Unification
schemes for AGN generally involve the following components: Kerr
black hole, a relativistic accretion disk, jets, broad line
clouds and a dusty torus.  For blazars the situation appears at
first sight to be a little different, and we wonder why that
might be.  There are two issues which should be considered in
connection with dusty tori in blazars.  Firstly, since there is
no direct evidence of any thermal emission from tori we should
consider what indirect evidence (at any wavelength) there is
about the existence of dust in the centres of host
galaxies. Secondly, we should consider what kind of torus
geometry would fit observations made at different wavelengths
best.

Since unification schemes propose that FR-I and BL Lacs are
related, FR-I sources being the mis-oriented counterparts of BL
Lacs \cite{UrryPadovani95}, we have searched the literature for
information about dust structures associated with blazars/BL
Lacs/FR-I.  We note that Blazejowski et al.~\cite{Blazejowski00}
required an external IR photon field from a torus scattered by
relativistic jet electrons, in addition to those in the SSC
model, in order to fit $\gamma$-ray spectra of OVV quasars
observed by EGRET. A key to the existence of tori in blazars
could come from establishing a link between the torus and the
BLR. For example, when there is no detection of direct optical
emission, the interpretation of spectro-polarimetric data on FR-I
objects with strong evidence for infrared obscuration
(\cite{Antonucci01} and references therein) could suggest the
existence of a BLR hidden by a thick torus. In this case, the
observed polarized broad lines would be the result of scattering
into the line of sight by free electrons in zones whose shape and
orientation is determined by the torus' inner geometry. Hence,
the detection of free electron scattering regions should be a
good diagnostic of torus geometry.  Falcke et al.~\cite{Falcke95}
proposed that the opening angle of the torus might play a
critical role in the FR-I and FR-II dichotomy -- a closed torus
covering a large fraction of $4 \pi$ steradians, as seen from the
black hole, would obscure the internal activity of FR-I objects.

de Koff et al.~\cite{Koff00} have discussed how the properties of
the dust depend upon the radio properties of the object, and
found that FR-II, which are powerful radio galaxies, have a
rather clumpy dust distribution, and they suggest that this
translates into a large opening angle of the torus.  A less
powerful radio jet (such as those in FR-I) would only slowly
disperse the dust, and this type of obscuration might contribute
to the deceleration of FR-I jets by entraining the material from
the torus.

We consider first the FR-I galaxy Centaurus A. Alexander et
al.~\cite{Alexander99} explain the IR spectrum of Cen~A using a
combined model of infrared emission from startbursts, cirrus
clouds and an AGN-type torus. A compact optically thick torus
with $r_{\rm out,torus}=3.6$ pc and an opening angle of
$\phi=30^{\circ}$ would account for the observed flux. Centaurus
A appears as a concentrated IR source in the H band
\cite{Bryant99} within a scale of tens of pc.  No optical
emission has been observed from the central source and, as Falcke
et al.~\cite{Falcke95} suggested, this source may have a closed
torus hiding the BLR which has not been detected even in
polarized flux.  Israel et al.~\cite{IsraelDishoeck90}, Rydbeck
et al.~\cite{Rydbeck93} and Turner et al.~\cite{Turner97}
concluded that there is a dense nuclear torus with diameter $<
230$ pc. It is also interesting that Bryant et
al.~\cite{Bryant99} show a clear picture of a warped dust lane
obscuring the nuclear source -- the radio jet being perpendicular
to a large-scale torus which seems to be inclined with respect to
the plane of the dust-lane. 

Turning now to other FR-I radio galaxies, 3C218 (Hydra A) also
shows evidence for a nuclear obscuration \cite{Sambruna00}.
3C270 is another FR-I radio galaxy for which a broad $H_{\alpha}$
emission line has been detected \cite{Barth99}. Ferrarese
et al.~\cite{Ferrarese99} proposed that its nucleus is surrounded by a dust
torus with a diameter of 120 pc and optical depth $\tau \approx
1$. An inclination angle of $20^{\circ}$ between the axis of the
torus and the line of sight would be in accordance with the
detection of un-obscured nuclear optical emission \cite{Chiaberge00}.

Possible evidence against tori in blazars comes from the FR-I
radio galaxy M87, which seems to have a very low IR flux which
cannot be explained with a standard torus model \cite{Perlman01}.
The torus in this source could be very diluted, and heated less as
a result of the extremely low accretion activity of M87
\cite{Reynolds96} such that its emission is dwarfed by the jet
emission.

At kiloparsec scales, starburst activity dominates the infrared
emission, but as one goes deeper into the nucleus of the galaxy,
de Koff et al.~\cite{Koff00} found that well organized dust
structures were present in FR-I galaxies in the 3CR catalog, and
this could indicate that the flow of matter towards the nucleus
is rather steady in FR-I, allowing the formation of distinct
torus features.  These structures tended to be sharply defined
small-scales disks with radii less than $\sim 2.5$ kpc.  The
detection of large-scale dusty features (sometimes shaped as bars
or dust-lanes) suggests that there could be an association
between a small-scale infrared torus, such as those in Seyfert
galaxies, and a large-scale torus sometimes identified with the
dust structure of the galaxy.  The structure of the nuclear torus
could then depend on the large-scale dust distribution --
well-organized kpc dust structures could extend inward towards
smaller scales (small-scale torus surrounded by large-scale
torus).  We define this as a symbiosis between the large and
small scale dusty features \cite{DoneaProtheroe02a}.

The above discussion leads us to the conclusion that dust could
be present everywhere, even in FR-I objects where the torus
cannot easily be detected. Since FR-I and BL Lacs are thought to
be similar (apart from orientation) and there is some evidence of
BLR and torus activity in FR-I, we postulate that the linkage
between BLR and tori in quasars discussed by Chiaberge et al.~\cite{Chiaberge00}
may apply also in blazars.  We shall calculate next the optical
depth for $\gamma$-$\gamma$ absorption in the infrared radiation
field of tori.

\begin{figure}
\begin{center}
\epsfig{file=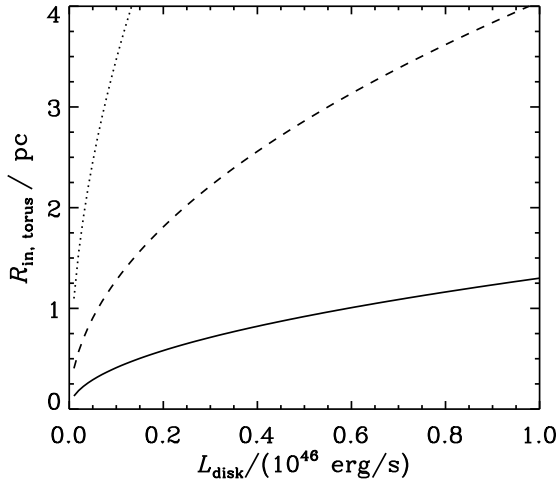,width=10.0cm}
\caption{$R_{\rm in, torus}$ vs. $L_{\rm disk}$ for
three different dust temperatures: $T=1500$~K
(solid curve), 1000~K (dashed curve), 700~K (dotted curve).}
\label{fig11}
\end{center}
\end{figure}

We know that the IR emission from the torus is strongly related
to the activity of the central object.  The inner radius depends
on $L_{\rm disk}$, and is given by the sublimation radius of the
dust \cite{Barvainis87}:
\begin{equation}
R_{\rm in, torus} \approx
T_{1500}^{-2.8} L_{\rm disk,46}^{1/2}~pc
\label{14}
\end{equation}
where $L_{\rm disk}=10^{46} L_{\rm disk,46}$~erg/s, and
$T_{1500}$ is the dust temperature in units of 1500~K (the dust
sublimation temperature is taken to be 1500~K).  The inner torus
radius is plotted in Fig.~\ref{fig11} versus disk luminosity for
three dust temperatures.

\begin{figure}
\begin{center}
\epsfig{file=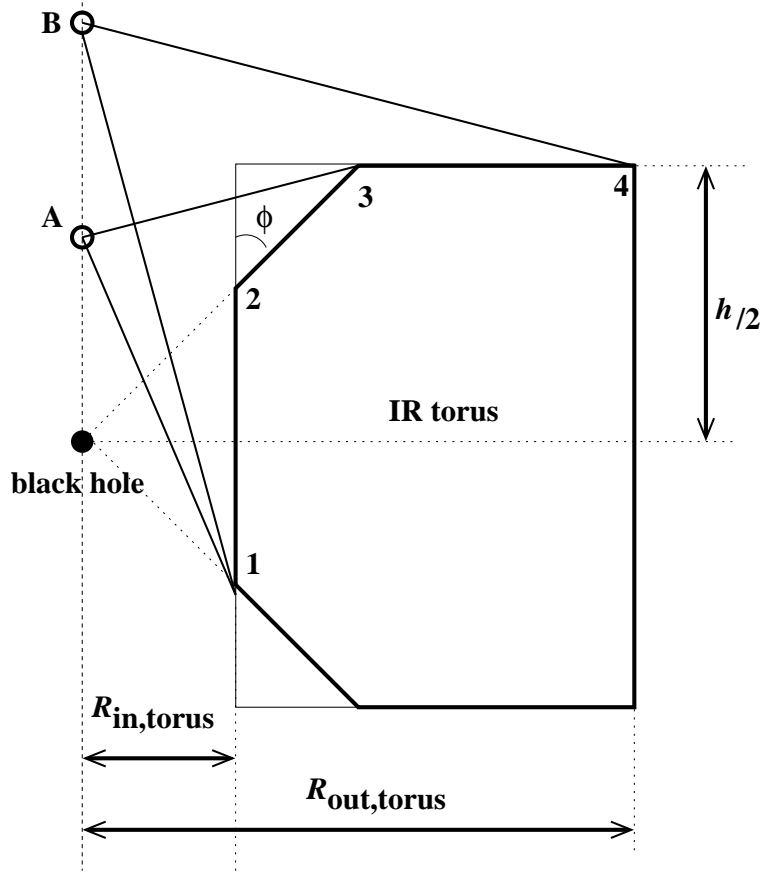,width=10.0cm} \caption{Torus cross-sections
      considered: rectangle, rectangle with edges cut away at
      angle $\phi$, the opening angle in this case (see text).
      $\gamma$-rays at point A interact with any photons emitted
      from the inner surface of the torus (between points 1 to
      3). Gamma-rays at point B (above the torus) also interact
      with any photons emitted from the upper surface (between 3
      to 4). }
\label{fig3}
\end{center}
\end{figure}

We start with a torus centred on the black hole, symmetric
about the jet axis, and having a rectangular cross section with
full height $h$, and inner and outer radii $R_{\rm in,torus}$ and
$R_{\rm out,torus}$. We shall discuss how $\tau_{\gamma\gamma}$
changes for different scales of tori, e.g., simulating an open
torus by having either a rectangular cross section or a
rectangular cross section with the inner edges cut away at angle
$\phi$ (see Fig.~\ref{fig3}).

As discussed earlier, in the context of the symbiosis between
jets and accretion disks, an increase in jet power could be at
the expense of disk luminosity.  A lower disk luminosity at UV
frequencies would reduce the heating of the inner surface of the
dusty torus, causing the inner radius of the torus, given by
Eq.~\ref{14}, to be small
(Fig.~\ref{fig11}). This implies that fat dusty tori ($R_{\rm
in,torus} \ll h$) could exist in AGN with very low central
activity such as blazars, and this would be in accordance with
the closed torus model discussed by Falcke et
al.~\cite{Falcke95}. On the other hand, for a ``cold torus'' with
$T<1000$~K, the inner radius would be far from the nucleus for a
given luminosity (see Fig.~\ref{fig11}).  Such inefficient
heating of the dust would probably describe best the torus in M87
which seems to be a peculiar object in that it has an extremely
low mass accretion rate.

Due to its size, the energy density of the infrared photons
remains fairly uniform inside the torus.  A flat torus (i.e., $R
\gg h$) could approximate the models proposed by Chiaberge et
al. \cite{Chiaberge99}; see the discussion from the previous
section and comments by Maiolino et al.~\cite{Maiolino01} about a
flattened BLR distributions. Although flat tori have not been
detected, and we are uncertain about the stability of such a
configuration of dust, we believe that it is an interesting
possibility.  For simplicity we model in this section a torus
with a single black body temperature such that the IR
intensity is constant within the solid angle subtended by the
torus.  We shall extend the work of Protheroe \& Biermann
\cite{ProtheroeBiermann97} for different torus geometries.
Fig.~\ref{fig3} illustrates the geometry for interaction, at
points A and B, of $\gamma$-rays with IR photons emitted from the
surfaces of the torus.

\begin{figure}
\begin{center}
\epsfig{file=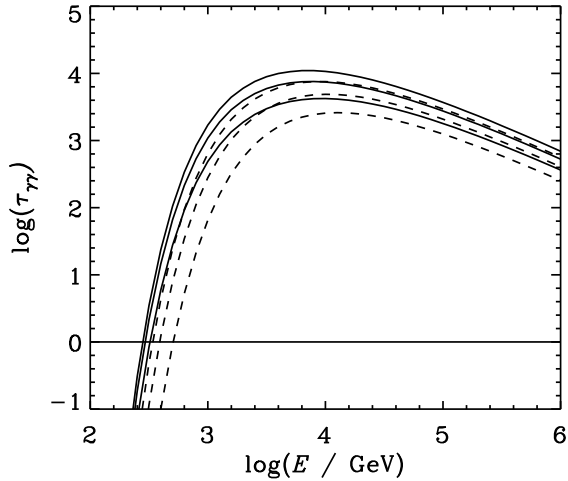,width=10.0cm} 
\caption{ Optical depths for $\gamma$-rays traveling along the
jet axis from $z=0$ to infinity in the torus IR radiation 
($T=1000$~K) for $r_{\rm out}=2$~pc and $r_{\rm in}=0.1$~pc (solid
curves) or $r_{\rm in}=1$~pc (dashed curves).  Results are show
for three torus heights: $h=1$~pc (lower curves), $h=2$~pc
(middle curves), $h=3$~pc (upper curves).}
\label{fig9}
\end{center}
\end{figure}

\begin{figure}
\begin{center}
\centerline{\epsfig{file=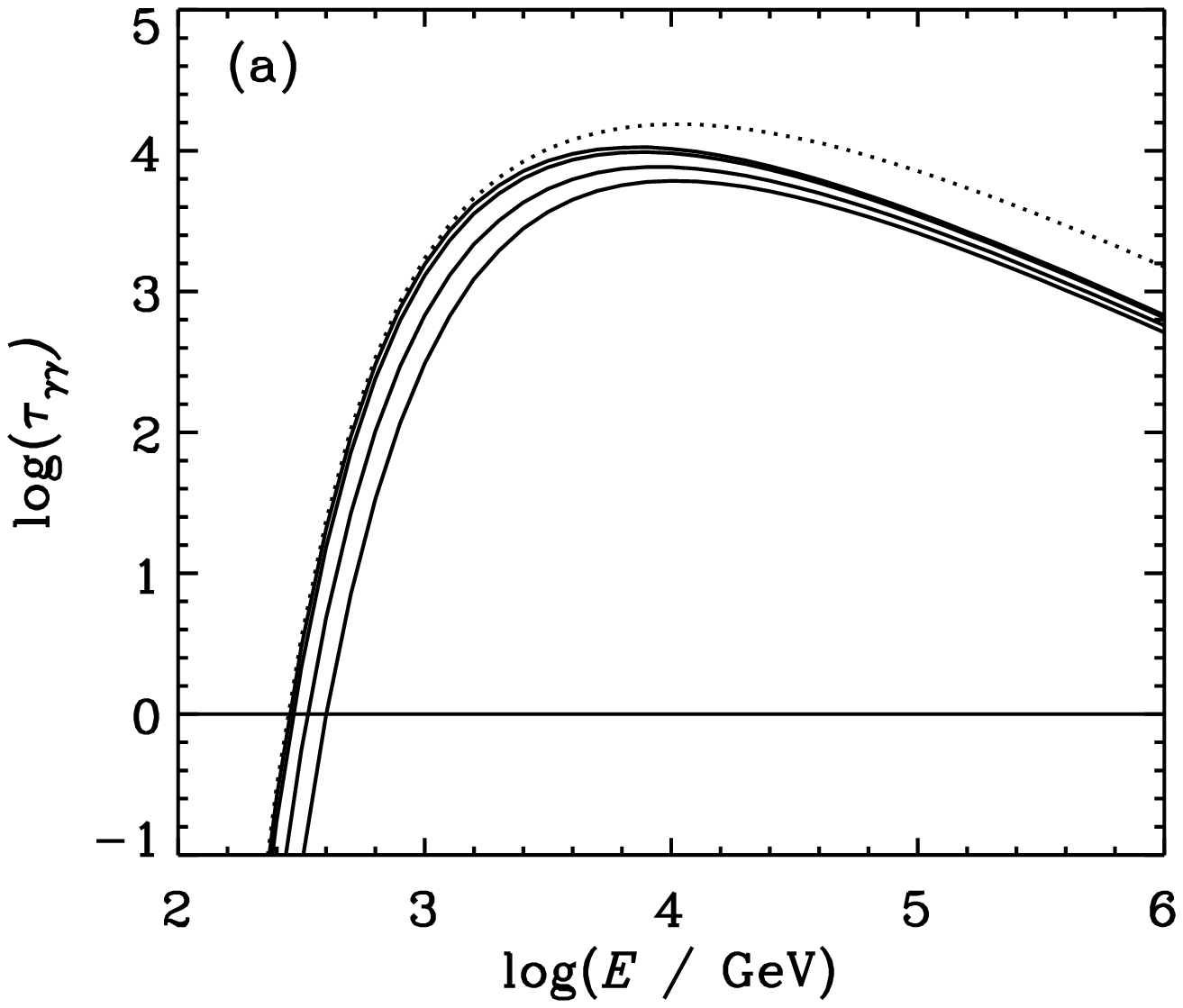,width=10.0cm}\hspace*{-5em}\epsfig{file=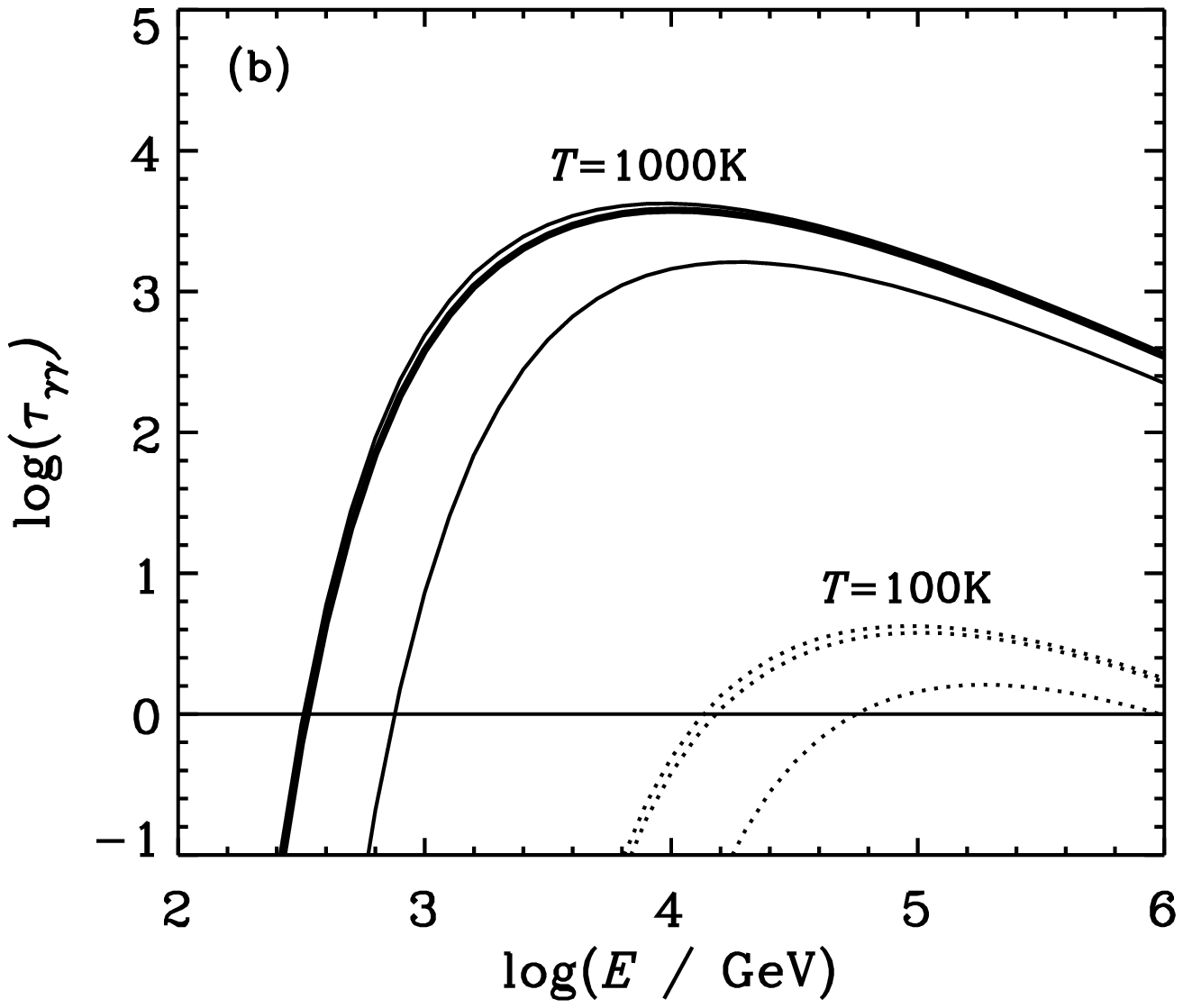,width=10.0cm}}
\caption{ Optical depths for $\gamma$-rays traveling along the
jet axis from $z=z_0$ to infinity in the torus IR radiation. (a)
$T=1000$~K, $z_0=0$, $r_{\rm in}=0.1$~pc, $r_{\rm out}=2$~pc,
$h=3$~pc and with the torus inner edge cut away at angle
$\phi=0^\circ$ (upper solid curve), $10^\circ$, $30^\circ$,
$50^\circ$ (lower solid curve).  Dotted curve is for $r_{\rm
out}=10$~pc and $\phi=0^\circ$.  (b) $r_{\rm in}=0.1$~pc, $r_{\rm
out}=2$~pc, $h=1$~pc, $\phi=0^\circ$, $T=1000$~K (solid curves)
or $T=100$~K (dotted curves), for $z_0=0.01$~pc (upper curves),
0.1~pc (middle curves), and 0.5~pc (lower curves)}
\label{fig9vary}
\end{center}
\end{figure}

Accretion rates are smaller in blazars than in quasars.  Since
the thickness of the torus is related to the accretion inflow and
disk luminosity (through radiation pressure effects)
\cite{Dopita98} one expects tori in FR-I to be thinner, diluted,
and/or less efficiently heated -- as is probably the case in M87.
Therefore, we have varied $h$ between $1$~pc and $3$~pc, $R_{\rm
in, torus}$ between $0.1$~pc and 1~pc, and $R_{\rm out, torus}$
between 2~pc and 10~pc. The optical depth from $z=0$ to infinity,
for $\gamma$-rays traveling along the jet axis, is shown in
Fig.~\ref{fig9} where we have compared the case of a torus with
$R_{\rm in, torus}=0.1$~pc (solid curves) with a torus having
$R_{\rm in, torus}=1$~pc (dashed curves) for three different
torus heights $h$.  We find that more $\gamma$-rays with energies
around $10^{2.5-3.5}$~GeV would be absorbed by $\gamma$-$\gamma$
interactions when the torus comes closer to the central source.
We also find that an open torus geometry ($\phi >0$) can
significantly modify $\tau_{\gamma\gamma}$ as shown in
Fig.~\ref{fig9vary}(a) where it is again seen to have the
greatest effect near the pair production threshold.  If the torus
has a large outer radius, then more IR photons from the upper
surface of the torus are available for interaction with
$\gamma$-rays along the jet. For example, for an extended torus
with $R_{\rm out, torus}=10$~pc the absorption of photons with
energy above 2~TeV is larger (compare dotted line with upper
solid curve in Fig.~\ref{fig9vary}a). As noted previously by
Protheroe and Biermann \cite{ProtheroeBiermann97}, we can see
that for photon energies above about several hundreds of GeV the
opacity is large enough such that no TeV $\gamma$-rays can emerge
if the source is near the centre of the torus.

Risaliti et al.~\cite{Risaliti99} have shown that a torus is
stable if the mass of the dust in the torus does not exceed the
dynamical mass flowing toward the center. This constrains the
outer torus radius, $R_{\rm out, torus}$, to have values less than
10~pc for a column density of dust along the line of sight that
intersects the obscuring medium of $N\approx 10^{24}$~cm$^{-2}$.
FR-I objects could have diluted tori with column densities of
$N\approx 10^{22}$~cm$^{-2}$, or lower, but still be optically
thick, and this would impose a upper limit of $\sim 100$~pc for
the torus' outer radius. However, Maiolino et al.~\cite{Maiolino98}
interpret their data as suggesting that the inner parts of tori
are much denser than the outer parts, and find a gradient in the
torus' covering factor that would give $R_{\rm out, torus}< 20$~pc
for Seyfert galaxies. The covering factor plays a crucial role in
these models, as well as the gas mass enclosed within the torus
\cite{Risaliti99}.  For simplicity, we take a value of 10~pc as
the maximum radius of the dust torus which could be relevant to our
analysis.  

In Fig.~\ref{fig9vary}(b) we show how the optical depth depends
on the distance of $\gamma$-ray emission region along the jet.  So
far, we have assumed that the torus radiates as a perfect black
body.  However, if the torus is patchy or not optically thick at
IR wavelengths we may have diluted black body radiation, $u_{\rm
IR} \approx \eta_{\rm IR} a T^4$~erg/cm$^3$, where $a$ is the
radiation constant.  All our curves would be multiplied by
$1/\eta_{\rm IR}$ in this case.  Alternatively, for a poor
heating mechanism related to a low luminosity of the accretion
disk the dust could be heated to lower temperatures, e.g.
$T\approx 100$~K, and this results in reducing the GeV to TeV
opacity considerably, with the pair production threshold
$\gamma$-ray energy increasing as $\sim 1/T$ as illustrated in
Fig.~\ref{fig9vary}(b) by the dotted curves. We note however,
that Blazejowski et al.~\cite{Blazejowski00} require $\eta_{IR}
\approx 0.3$ to fit $\gamma$-ray spectra of OVV quasars with an
external inverse Compton model, and this still allows for
significant TeV $\gamma$-ray absorption in this type of
blazar. Extremely low values of $\eta_{IR}$ may apply in objects
such as M87.


\section{Constraints on the position of the $\gamma$-ray 
emitting region and conclusions}

The ADJ model gives a simplified approach to
the symbiosis between disks and jets. It could explain the low
central activity in some blazars, and for very low mass accretion
rates an ADJ  could transform into an EDAF or an ADAF. The strong
variability observed in blazars from radio to $\gamma$-ray
frequencies could be a consequence of the feeding mechanism of
the jet at the boundary layer. A local perturbation produced in
the boundary layer, such as a rapid variation of $R_{\rm jet}$ or
$\dot M$, would propagate through the nozzle and along the jet causing
transient phenomena.  As the jet extracts more energy from the disk, 
this could leave less energy to be dissipated in the disk as
radiation.  The jet/disk symbiosis then mainly modifies the energetics
in the central parsec of AGN: increasing the power of the jet
could weaken the disk which therefore would not efficiently
ionize the BLR clouds or efficiently heat the dust in the torus,
and so would affect important target photon fields for
interactions of $\gamma$-rays (and also accelerated electrons and
protons) along the jet.

\begin{figure}
\begin{center}
\centerline{\epsfig{file=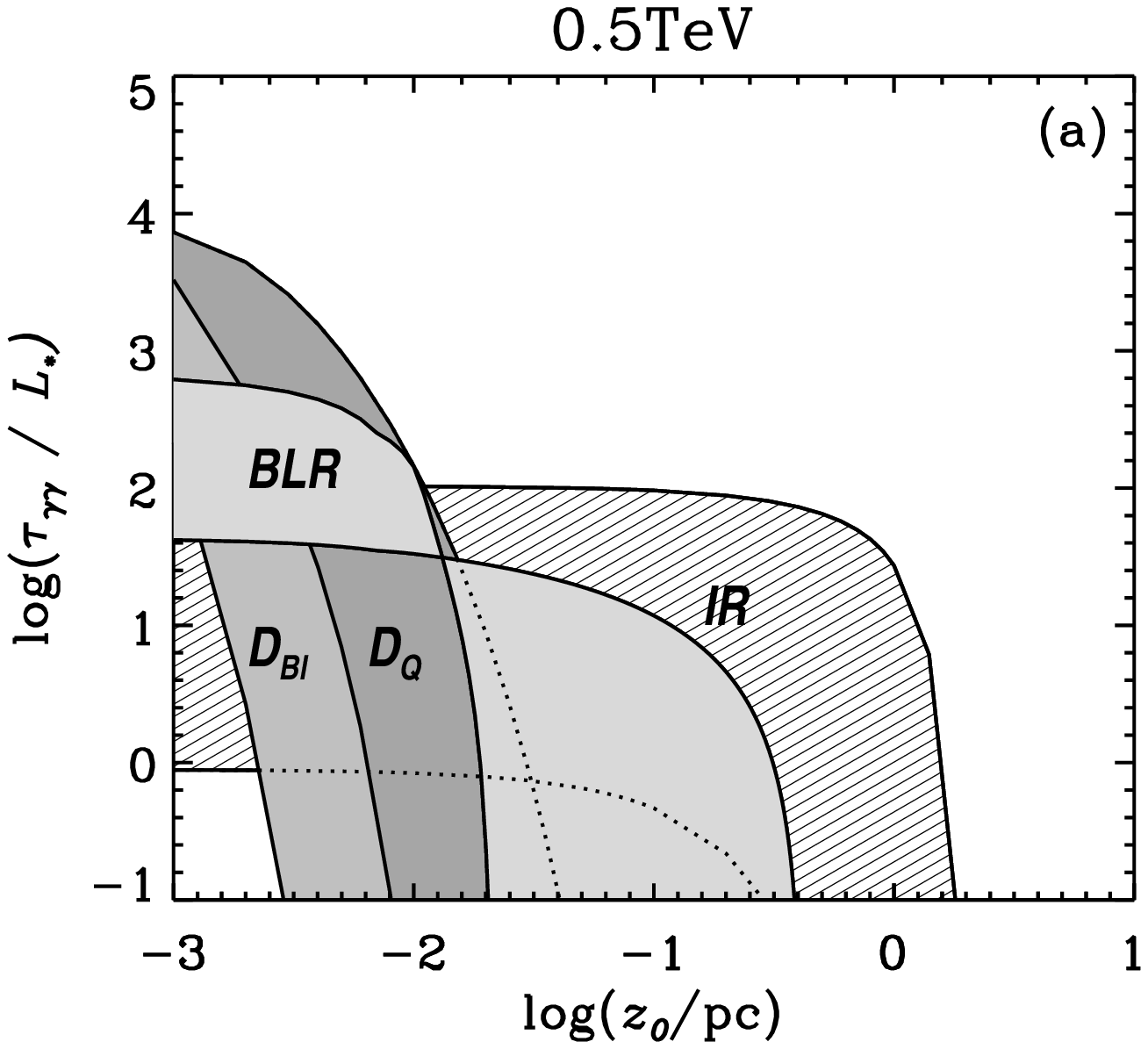,width=10.0cm}\hspace*{-5em}\epsfig{file=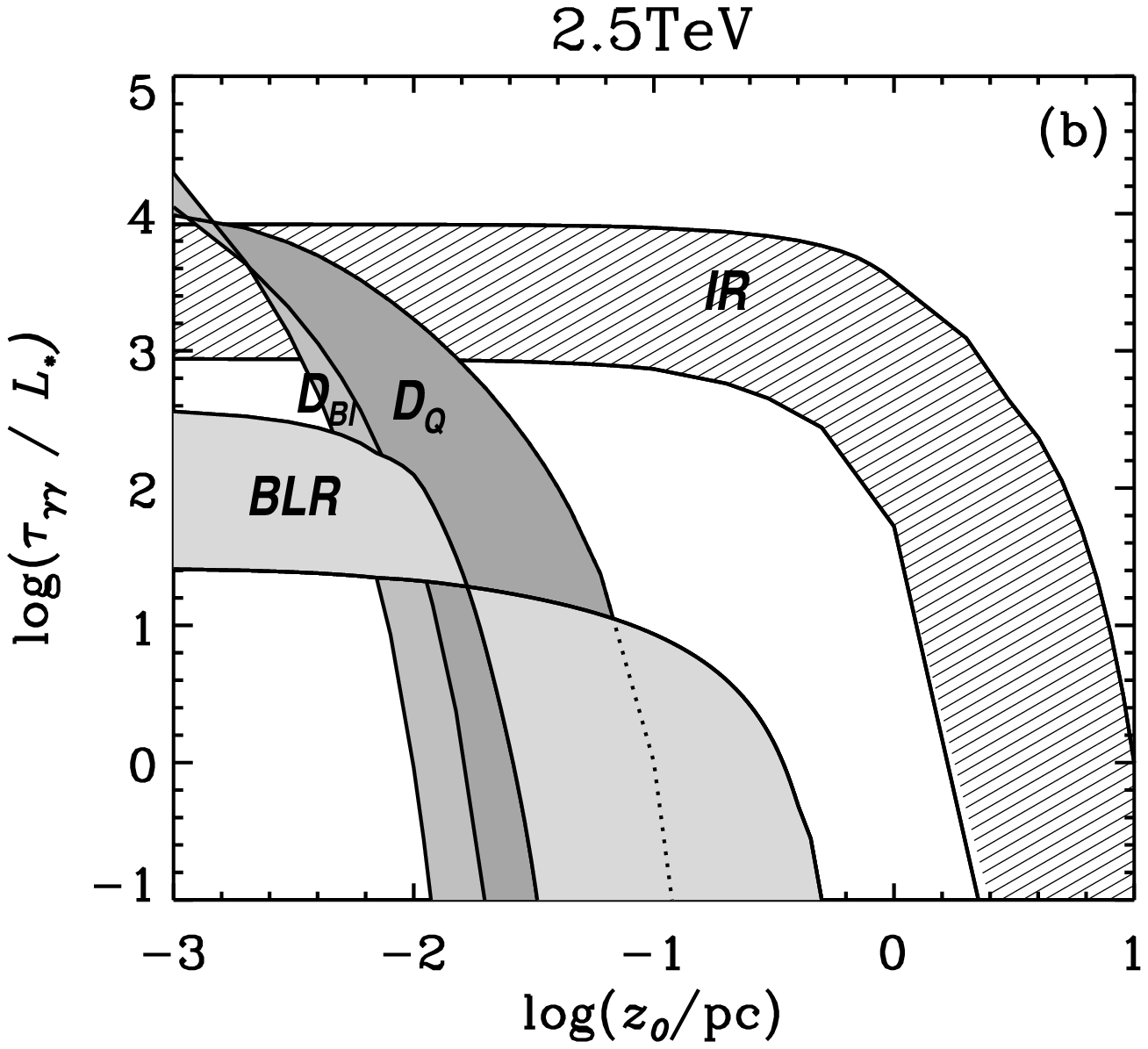,width=10.0cm}}
\centerline{\epsfig{file=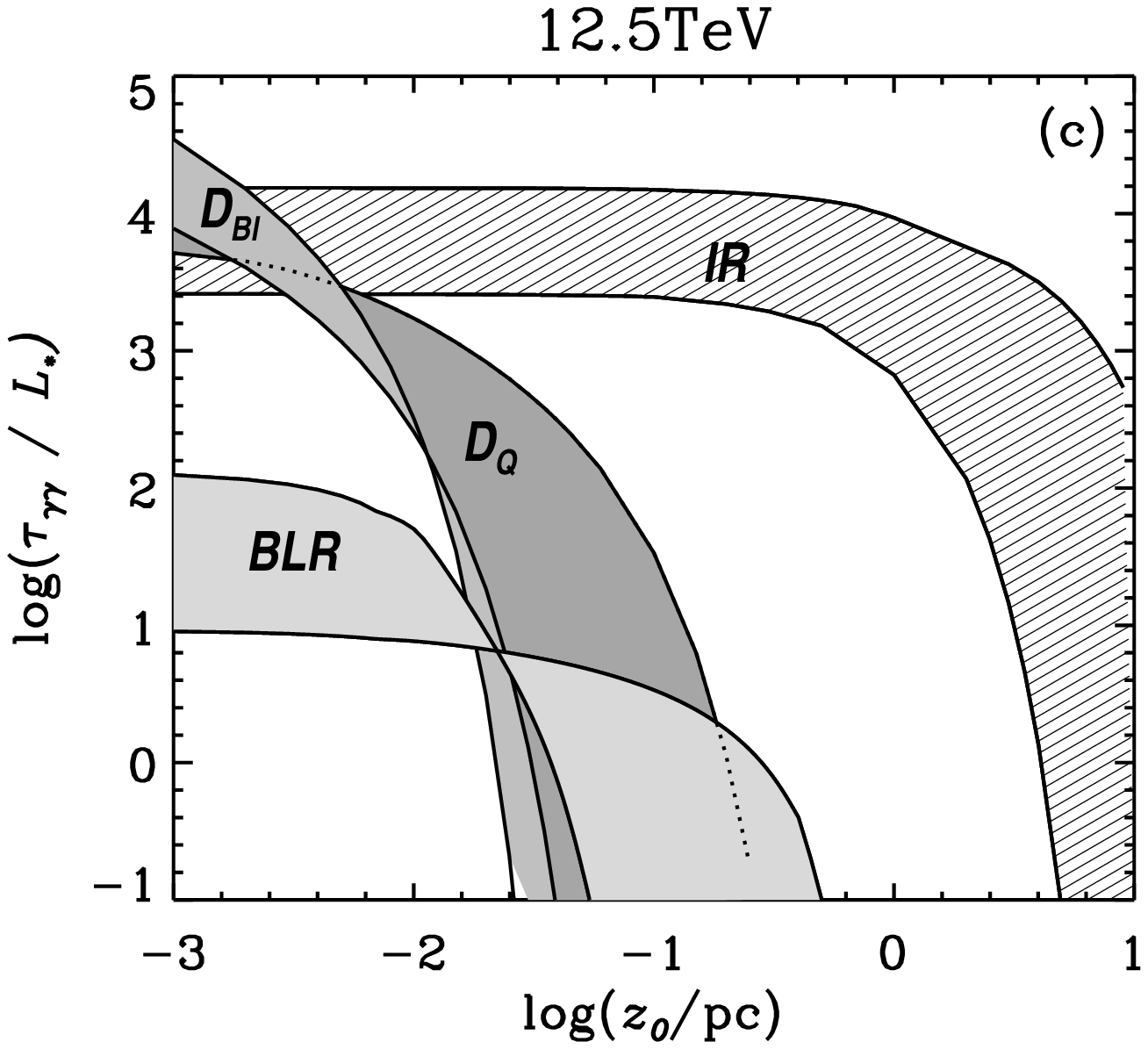,width=10.0cm}}
\caption{Photon-photon optical depth, divided by $L*$, vs.\
height of the emission region $z_0$ above the disk for
interaction with BLR photons (``BLR'', $L*=L_{\rm
BLR}/10^{44}$~erg s$^{-1}$), IR photons (``IR'', $L*\equiv
\eta_{\rm IR}$) and accretion disk photons in blazars (``D$_{\rm
Bl}$'', $L*=L_{\rm disk}/10^{46}$~erg s$^{-1}$) and quasars
(``D$\rm _Q$'', $L*=L_{\rm disk}/10^{46}$~erg s$^{-1}$).  The
ranges of model parameters used for the three radiation fields
are: accretion disk radiation -- as in Fig.~2b; BLR -- $\tau_{\rm
BLR}=0.01$, $R_{\rm BLR}=0.01$~pc (upper boundary on left is as
in Fig.~6a), $R_{\rm in,BLR}=0.01$~pc, $R_{\rm out,BLR}=0.4$~pc
(lower boundary on left is as in Fig.~6b); torus -- $T=1000$~K,
$\phi=0$, upper curve ($h=3$~pc, $r_{\rm in}=0.1$~pc, $r_{\rm
out}=10$~pc), lower curve ($h=1$~pc, $r_{\rm in}=1$~pc, $r_{\rm
out}=2$~pc).}
\label{fig7}
\end{center}
\end{figure}

Fig.~\ref{fig7} summarizes our results in a way that makes it
easy to see the relative importance of the three radiation fields
to the absorption of high energy $\gamma$-rays.  We show the optical
depth as a function of height of the emission region above the
plane of the disk for three $\gamma$-ray energies spanning the range
currently explored with optical Cherenkov telescopes: (a) 0.5~TeV
which is close to the threshold of present telescopes, (b) 2.5~TeV,
(c) 12.5~TeV typically observed during flaring activity in nearby
blazars.  The shaded areas give the range in optical depth for
the models discussed earlier, but normalized for an accretion
disk luminosity $L_{\rm disk}=10^{46}$~erg s$^{-1}$, a broad line
region luminosity $L_{\rm BLR}=10^{44}$~erg s$^{-1}$, and torus
IR dilution factor $\eta_{\rm IR}=1$.  We see that at all three
energies, all three radiation fields can play a very important
role.  We shall consider each radiation field in turn, first for
quasars and then for blazars.

In quasars, with typically $L_{\rm disk} \ge 10^{46}$~erg
s$^{-1}$ the accretion disk radiation will surely cut off
$\gamma$-rays above 0.5~TeV if the emission region is below $\sim
0.005$--0.03~pc (depending on $\dot{m}$).  At 2.5~TeV and
12.5~TeV the corresponding minimum emission region heights to
avoid absorption in quasars are $\sim 0.02$--0.08~pc and $\sim
0.03$--0.2~pc.  The BLR is active in quasars, and is likely to
have $L_{\rm BLR} \ge 10^{44}$~erg s$^{-1}$ such that the
emission region would have to be above 0.03--0.3~pc (depending on
BLR cloud distribution) at all three energies.  For a luminous
disk as in quasars, one would expect an active torus to cut off
TeV energy $\gamma$-rays unless the emission region were well
above the torus: $z_0>0$--2~pc (0.5~TeV, note
$\tau_{\gamma\gamma}^{\rm IR}$ can be small at the energy),
$z_0>2$--10~pc (2.5~TeV), and $z_0>4$--30~pc (12.5~TeV),
depending on torus geometry.  All known quasars are at a distance
such that multi-TeV $\gamma$-rays are expected to be absorbed in
the infrared background radiation during their propagation to
Earth.  However, observations at energies below 1~TeV are
possible with current atmospheric Cherenkov telescopes, and
multi-GeV observations will be made with GLAST~\cite{GLAST} within a few years.
Future measurements of the shape of the $\gamma$-ray spectrum of
quasars at the cut-off might have something to say about which
radiation field is responsible for the cut-off, and hence the
location of the emission region, although this is likely to be
very difficult (compare the optical depth vs.\ energy in the disk
radiation, BLR radiation and torus radiation in Figs.~2, 6 and
11).

The situation is somewhat different in blazars.  Here, we can
have significantly less powerful accretion disks than in quasars.
For example, to avoid attenuation of $\gamma$-rays in the disk
radiation for a disk luminosity of $10^{44}$--$10^{46}$~erg
s$^{-1}$ one would need to be above $\sim 0.001$--0.005~pc
(0.5~TeV), and above $\sim 0.005$--0.03~pc ($> 2.5$~TeV),
depending on $\dot{m}$.  Note that at all energies the optical
depth in the BLR field becomes less than unity if the BLR
luminosity is lower than $10^{42}$~erg s$^{-1}$.  However, for
$L_{\rm BLR}=10^{42}$--$10^{44}$~erg s$^{-1}$ the $\gamma$-ray
emission region would need to be above 0.03--0.3~pc (depending on
BLR cloud distribution), as for quasars, to avoid attenuation.  In
the case of the torus radiation, if blazars have active tori,
they could present a serious problem for the escape of TeV
$\gamma$-rays from emission regions which are not well above the
torus (as in the case of quasars).  However, because of the low
disk luminosity, the torus' inner radius would probably be small
($\sim 0.1$~pc for $L_{\rm disk} \ge 10^{44}$~erg s$^{-1}$, see
Eq.~11), and the torus would be likely to be small, generally,
and the optical depth would probably be at or below the lower
boundary of that shown.  Inefficient heating could also result in
dilution of the torus radiation ($\eta_{\rm IR} <1$) or in a
lower temperature, and consequently lower optical depths (see
e.g.\ Fig.~11b).

We have demonstrated that the external radiation fields present
within the parsec scale of AGN may create problems for the escape
of high energy $\gamma$-rays. The effect is dramatic for quasars
where the densities of external radiation fields are much higher
than in blazars. The symbiosis between the accretion disk and
jets could result in lower radiation fields around the base of
the jet during flaring activity if the jet takes a higher
fraction of the matter entering the disk, and this could minimize
the dramatic effect of the $\gamma$-ray absorption.  However,
although heated by the disk, the torus IR field would not react
rapidly to changes in disk luminosity because of its larger
scale, and so the escape of TeV $\gamma$-rays would require the
emission region to be located at parsec distances from the black
hole, far above the torus unless the torus is very inefficiently
heated although even in this case one can expect some
attenuation.  This is in accordance with conclusion of Protheroe
and Biermann \cite{ProtheroeBiermann97}, and also the recent work
of Jorstad et al.~\cite{Jorstad01} who concluded that the temporal
connection between the radio and $\gamma$-ray variations suggests
that the $\gamma$-rays appear to originate far away from the
radio core.

In conclusion, the origin of the 100~GeV-1000~GeV $\gamma$-ray
photons must be above the inner part of BLR region for any
quasar.  Blazars usually have low central luminosities, and this
translates into a having a low BLR luminosity, and possibly a
closed torus.  Mrk 421 seems to have a BLR with an extremely low
luminosity $L_{\rm BLR} \approx 1.5\times 10^{40}$~erg/s
\cite{Morganti92} and so could be transparent to TeV $\gamma$-rays.
However, TeV $\gamma$-rays would need to be produced far above
the torus unless the torus is very inefficiently heated.  Farther
along the jet external radiation fields become less important.
In this case the radiation produced inside the emission region in
the jet (e.g.\ synchrotron emission) becomes the dominant
radiation field for photon-photon pair production by $\gamma$-rays
produced in the jet.

\section*{Acknowledgments}
We thank Peter Biermann and Geoff Bicknell for useful discussions.  This work
was supported by an Australian Research Council grant to RJP.

\end{document}